# Variable Augmented Network for Invertible Modality Synthesis-Fusion

Yuhao Wang, Ruirui Liu, Zihao Li, Cailian Yang, Qiegen Liu

*Abstract*—As an effective way to integrate the information contained in multiple medical images under different modalities, medical image synthesis and fusion have emerged in various clinical applications such as disease diagnosis and treatment planning. In this paper, an invertible and variable augmented network (iVAN) is proposed for medical image synthesis and fusion. In iVAN, the channel number of the network input and output is the same through variable augmentation technology, and data relevance is enhanced, which is conducive to the generation of characterization information. Meanwhile, the invertible network is used to achieve the bidirectional inference processes. Due to the invertible and variable augmentation schemes, iVAN can not only be applied to the mappings of multi-input to one-output and multi-input to multi-output, but also be applied to one-input to multi-output. Experimental results demonstrated that the proposed method can obtain competitive or superior performance in comparison to representative medical image synthesis and fusion methods.

*Index Terms*—Image synthesis, image fusion, invertible network, variable augmentation.[1]

## I. INTRODUCTION

As well known, medical imaging plays an important role in a variety of clinical applications. Due to the limitation of the imaging mechanism, medical images from a single modality usually cannot provide sufficient information to meet the requirements of complex diagnoses [1]. For instance, computerized tomography (CT) [2] images can provide a clear visualization of dense structures like bones and implants, yet it is not good at presenting the soft tissues. Magnetic resonance imaging (MR) [3] images can provide high-resolution detailed information of soft tissues, while it is also prone to introduce artifacts when taking photos of bone structures [4]. Functional information of blood flow and metabolic changes can be reflected by positron emission tomography (PET) images, but the spatial resolution is usually very low. Multi-modal medical image synthesis and fusion provide effective candidate techniques to address this issue. The synthesis of medical images from one modality to another is an intensity transformation between two images acquired from different medical devices. The medical image fusion is to obtain complementary information from multiple modalities, such that richer medical information and more details can be obtained. At present, there are few methods to tackle image synthesis and fusion simultaneously. For the ease of readability, we introduce some works on medical image synthesis and fusion as below:

*Medical Image Synthesis.* In recent years, a variety of medical image synthesis methods have been proposed [5]-[25]. Due to the difference in imaging mechanism, the intensities of different source images at the same location often vary significantly. Image synthesis can realize modality transformation and make full use of the characteristics of each modality. Traditional medical image synthesis methods are often regarded as patch-based regression tasks [5], [6], which take a patch of an image or volume from one modality to predict the intensity of a corresponding patch in a target image. Next, a regression forest is utilized to regress target-modality patches from these given-modality patches. In addition to the patch-based regression models, there has also been rapid development in sparse representation for medical image synthesis [7], [8]. In [8], Huang *et al.* proposed a weakly supervised joint convolutional sparse coding to simultaneously solve the problems of super-resolution and cross-modality image synthesis. Another popular type of medical image synthesis method is the atlas-based model [9]. These methods [10], [11] adopt the paired image atlases from the source modality and target modality to calculate the atlas-to-image transformation in the source modality, which is then applied for synthesizing target-modality-like images from their corresponding target modality atlases.

In the past few years, deep learning methodology has been widely applied in medical image synthesis, and has reported good performance [12]-[14]. Almost existing approaches can be divided into three categories: Early CNN or FCN learning approaches, auto-encoders (AE), and generative adversarial network (GAN). Dong *et al.* [15] proposed an end-to-end mapping between low/high-resolution images using CNNs. Nevertheless, there is a potential vulnerability of the CNN-based synthesis methods that it is easy to ignore context information when synthesizing images. In addition, GAN has been widely applied in the field of image synthesis [12], [16]-[18]. Dar *et al.* [19] proposed a new approach for multi-contrast MRI synthesis based on conditional GAN, which preserves high-frequency details via an adversarial loss. More recently, some variants or hybrid approaches that taking advantage of AE and GAN have been developed [22]-[25]. For instance, Sharma *et al.* [20] and Zhou et al. [21] proposed variants of GAN to leverage implicit condition in multi-modal to realize many-to-one mapping. Chartsias *et al.* [22] applied a multi-input multi-output FCN to MRI synthesis. The model first learns to embed all input modalities into a shared modality-invariant latent space. Then, these latent representations are combined into a single fused representation. Meanwhile, Liu *et al.* [23] proposed a method based on GAN, using different MR images as the input of GAN, and extracted multi-scale feature maps from

This work was supported by National Natural Science Foundation of China (61871206, 61601450).

Y. Wang, R. Liu, Z. Li, C. Yang, and Q. Liu are with the Department of Electronic Information Engineering, Nanchang University, Nanchang 330031, China. ({liurui, lizihao91, yangcailian}@email.ncu.edu.cn, {wangyuhao, liuqiegen}@ncu.edu.cn).

the input images through encoding for image synthesis. Sun *et al.* [24] proposed a flow-based generative model with AE-like network architecture for MRI-to-PET image generation. Although invertible blocks were involved in [25], the AE-like network architecture makes it not to be compact.

*Medical Image Fusion.* Multi-modal medical image fusion aims at combining the complementary information contained in different source images by generating a composite image for visualization [26]. Mainstream medical image fusion methods include decomposition-based and learning-based image fusion methods [27]-[30]. Liu *et al.* [31] proposed an integrated sparse representation and multi-scale transform-based medical image fusion framework. Zhu *et al.* [32] proposed a non-subsampled contourlet transform-based multi-modal decomposition method for medical images. These methods via multi-scale analysis would have the ability to select the frequencies in both space and time. Multi-scale transformation methods can not completely extract the effective information from the base layers of images, hence the fusion effect is not ideal in some situations.

Learning-based methods have been widely used in medical image fusion. Zhu *et al.* [33] designed a novel dictionary learning-based image fusion approach for multi-modal images. Due to the use of dictionary learning, this approach achieved high accuracy. But it was computationally exhaustive in essence. At present, CNN plays an increasingly important role in medical image fusion. Xia *et al.* [34] integrated multi-scale transform and CNN into a multi-modal medical image fusion framework, which uses the deep stacked neural network to divide source images into high and low frequency components to do corresponding image fusion. Liu *et al.* [35] proposed a CNN-based multi-modal medical image fusion algorithm, which applied image pyramids to the medical image fusion process in a multi-scale manner. CNN can fuse the images efficiently. However, it is a computationally extensive approach, and sometimes also not provides promising results if the images are very similar to each other. Ma *et al.* [36] designed a dual-discriminator conditional GAN to obtain multi-modal fused images. It obtained a real-like fused image by using the content loss to dupe both discriminators. Two discriminators were also considered intentionally to differentiate the composition variations between the fused and source images, respectively. Meanwhile, Wang *et al.* [37] developed a 3D auto-context-based locality adaptive multi-modal GAN to obtain efficient multi-modal fused images. A non-unified kernel was also used along with the adaptive approach for multi-modal fusion.

*Contribution.* In this study, inspired by the appealing work in [38] for decolorization and multi-exposure fusion, an invertible and variable augmented network (iVAN) is proposed for both invertible medical image synthesis and fusion. At first, we treat the unimodal problem and the multi-modal problem as a single problem that can be handled. More precisely, we use the core idea of increasing auxiliary variables to achieve a many-to-many mapping, so as to reduce the two problems into one. Besides, by incorporating the variable augmentation ideology into the invertible neural network, the resulting iVAN network achieves excellent performance. Since both the variables in network input and output participate in the design of the loss function, the network architecture as well as the loss function can be very simple, compared to the recent state-of-the-art networks.

The main contributions of this work are summarized as follows:

- **Invertible Synthesis and Fusion:** To the best of our knowledge, this is the first work that introduces the invertible network to tackle the two problems of medical image synthesis and fusion in a unified framework. Subsequently, an invertible modality synthesis-fusion system is formed.
- **Variable Augmentation Technology:** To enable the invertible network to be effective, variable augmented technology is employed to make the network input and output to be the same dimension, thereby treating image synthesis and fusion as a single problem.
- **Flexibility and Versatile Applications:** Due to the invertible and variable augmentation schemes, the present network can not only be applied to the mappings of multi-input to one-output and multi-input to multi-output, but also be applied to one-input to multi-output.

The rest of this paper is organized as follows. Section II introduces the related formulas and invertible neural networks. Section III presents the unified framework, network architecture, training objectives, and adversarial learning interpretation of the proposed network. Experimental results are provided in Section IV. The discussion of loss function and network channel is stated in Section V. Finally, Section VI concludes the paper.

## II. PRELIMINARY

Recently, many researchers have developed networks that can not only achieve unimodal synthesis, but also multi-model synthesis [22], [21], [39]. In this study, the present iVAN is still dedicated to these tasks. Additionally, we also attribute the fusion problem to the framework of image synthesis.

### A. Formulation of Synthesis and Fusion

*Unimodal Synthesis.* The goal of unimodal synthesis is to learn a one-to-one mapping. Mathematically, the medical image synthesis problem can be expressed as:

$$\tilde{y} = f(x) \quad (1)$$

where $x$ is the source modality image and $\tilde{y}$ is the synthesized modality image. It is worth noting that early studies on image synthesis mainly focus on unimodal synthesis [15], [16], [41]. Since multi-modal medical data can provide a wealth of diagnostic information, the demand for processing and analysis of multi-modal medical images is increasing [20]-[22].

*Multi-modal Synthesis and Fusion.* Multi-modal synthesis and fusion involve the same formulation, while exhibit different underlying tasks. Concretely, the former uses the complementary information of different modalities to obtain a high-quality target modal image. On the other hand, the latter is to acquire the fusion image containing different modality information. Both of them fall into the category of multi-input one-output mapping formulation, i.e.,

$$\tilde{y} = f(x_1, x_2, \dots, x_n) \quad (2)$$

where $x_1$, $x_2$, $x_n$ are different modality images, and $\tilde{y}$ is the synthetic or fused image. We treat each input modality as a channel, and then stack them up as the network input of iVAN.

### B. Invertible Neural Networks

Traditional neural networks generally consist of multiple layers of neurons and weighted connections between the layers [40]. Activation functions of the neurons in the hidden layers enable the neural networks to capture the nonlinear relation between the input and output data. These networks focus on forwarding prediction processes. Due to the limitation of the network structure, they do not have inverse inference capability. To handle the inverse process, flow-based generative models [24], [25], [42], were proposed. Unfortunately, the actual computation of these flow-based inversion was too costly to be practical.

Invertible neural networks (INNs) are the kind of methods that learn information-lossless mapping, which has also been extensively studied in flow-based generative models. INNs usually consist of a sequence of invertible layers, such as affine coupling layers [25], invertible 1×1 convolutional layers [43], and actnorm layers [44]. The general diagram of INN is illustrated in Fig. 1. INNs learn the mapping as $y = f(x)$, which is fully invertible as $x = f^{-1}(y)$. For an image pair $(x_i, y_i)$, with only one invertible network $f$, inputting the image $x_i$ into the network yields the output image $y_i$ and vice versa. Thus, the information is fully preserved during both the forward and reverse transformations. Due to the properties such as reversibility, memory savings, and simple design, INN is selected as the footstone of the propose medical synthesis and fusion method in this paper.

A requirement in INN is that the channel dimension of network input and output should be the same. In the next section, the variable augmentation technique will be introduced in INN to enable its diverse applications in image synthesis and fusion.

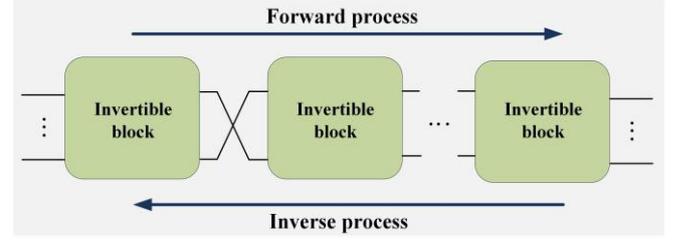

Fig. 1. Structure of INN consisting of reversible blocks.

## III. PROPOSED iVAN MODEL

### A. Unified Framework of Invertible Synthesis and Fusion

We conduct our research from four aspects shown in Fig. 2. In the experiment, we consider three different image synthesis scenes. The one-input to one-output mapping in Fig. 2 (a) is the earliest form of image synthesis task. That is, for a given modality $x$ as the network input, the output data $\tilde{y}$ is of another modality. Based on the principle of "more modalities provide more information", several studies have begun to investigate the multi-modal data synthesis. After that, MM Synthesis [22], Hi-Net [21], and MM-GAN [20] further applied a number of approaches for multi-input to one-output synthesis.

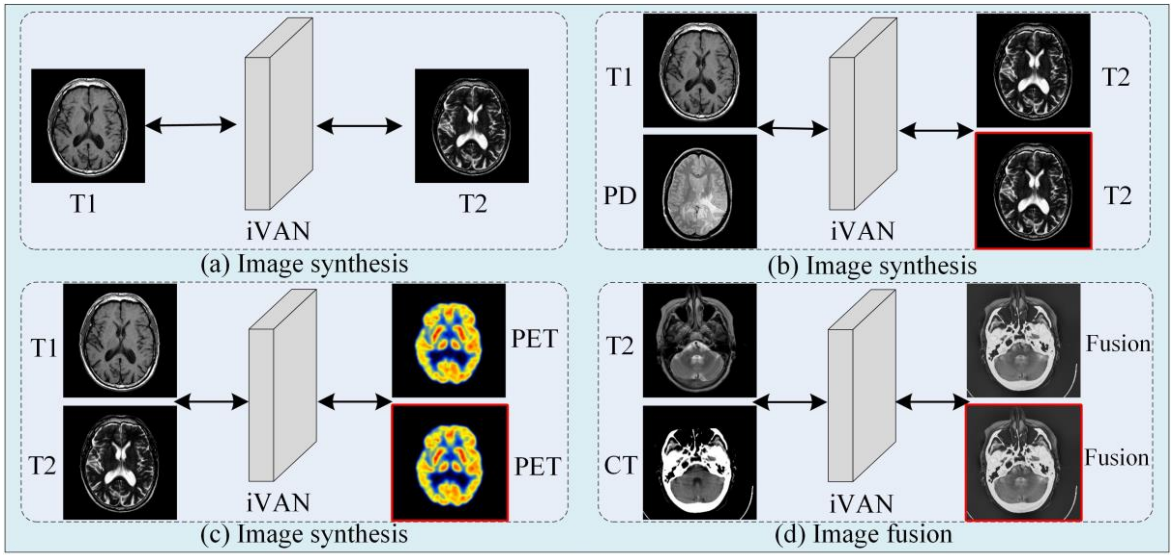

Fig. 2. Visual illustration of the invertible medical image synthesis and fusion in variable augmentation manner. Notice that the present network can tackle the synthesis cases of one-input to one-output, multi-input to one-output, and the fusion case of multi-input to one-output. The image in the red-square box is the variable augmented image.

Fig. 2(b) and (c) are the instances of multi-modal synthesis. Since the structural similarity between the network output and input in Fig. 2(c) is worse than that in Fig. 2(b), the synthesis in the latter case is more challenging. Due to the inconsistency between the input and output, there arise many difficulties. To achieve a proper multi-modal synthesis, one critical challenge is effectively fusing the various inputs. In MM Synthesis [22], Hi-Net [21], and MM-GAN [20], they employ encoders to embed all input modalities into the shared modal invariant latent space. Then, these latent representations are combined into a single fused representation. Finally, the representation is converted to the target output modality by the learned decoder to perform cross-modal synthesis tasks. The above methods are not only complicated in network structure, but also not have inverse capability. In this study, iVAN is used to solve image synthesis and fusion issues. Not only the synthesis scenarios under one-to-one, multi-to-one, multi-to-multi can be applied, but also the case of one-to-multi can be attained.

In iVAN, under the spiritual idea of variable augmentation [38], we add dummy variables and copy them in the output, so that the network input and output have the same channel dimensions. More specifically, in Fig. 2(b), $X = (x_{T1}, x_{PD})$ is used as the input of the network and the network output is $Y = (\tilde{y}_{T2}, \tilde{y}_{T2})$. Since PET is an RGB-channel image, in Fig. 2(c), $X = (x_{T1}, x_{T2}, x_{T1}, x_{T2}, x_{T1}, x_{T2})$ is used as the input of the network and the network output is $Y = (\tilde{y}_{PET}, \tilde{y}_{PET})$. For a given image, the invertible network we proposed can achieve multi-input to multi-output mapping. At the same time, we extend the iVAN in Fig. 2(d) to achieve a wider range of applications.

We take advantage of the inherent invertibility of normalizing-flow-based models, and apply iVAN to achieve invertible image synthesis and fusion. The utmost advantage of iVAN is that it does not disturb the modality-specific structure and only learns the underlying correlations among the modalities. What is more, it provides an invertible connection between the original image and the target image. Specifically, the model is designed with the composition of a stack of affine coupling layers and utilizes the invertible 1×1 convolution as the learnable permutation function between the coupling layers. The visualization of the training process in iVAN is shown in Fig. 3. Specifically, the forward process of iVAN produces the synthesis images, and the inverse process aims at recovering the original images. A combination of the forward loss and backward loss is used to optimize iVAN. iVAN can achieve the unity of fusion and invertible fusion (i.e., signal/image separation).

In the experiments, we use the sliced 2D images for training, slice the image, and normalize it to obtain more data information to increase the matching of the data. Then, we perform data augmentation processing on the input data to make the input data more complete and use the augmented data to enter the invertible network model for learning. In the training process, for image synthesis, we employ the original modality image as the network input, and the target modality image as the label; For images fusion, we use different modality images as the network input, and the target fused image as the label.

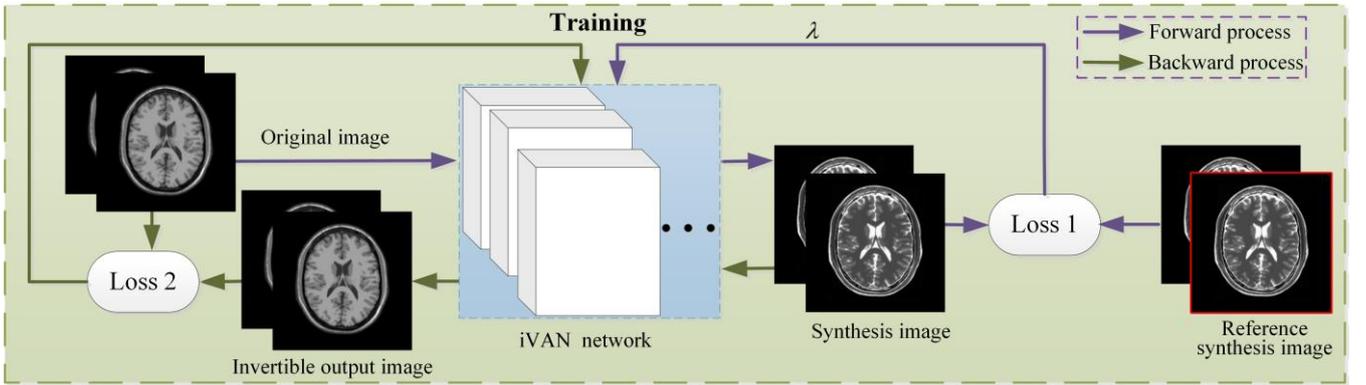

Fig. 3. The training pipeline of iVAN. During the training process, the network input and output are under the same dimension. The image in the red-square box is the variable augmented image. For the convenience of readers, the detailed architecture of iVAN is described in Section III. B. On the other hand, the description of the two loss functions (i.e., Loss 1 and Loss 2) in the training objectives is provided in Section III. C.

### B. Network Architecture of iVAN

Our goal is to find a bijective function which can map the data point from input data space $X$ to output data space $Y$. To achieve this, classical neural networks need two separate networks to approximate $X \to Y$ and $Y \to X$ mappings respectively, which leads to inaccurate bijective mapping and may accumulate the error of one mapping into the other. We take an alternative method and use the affine coupling layers in [45], [25] to enable invertibility of one single network. We design our iVAN with the composition of a stack of invertible and tractable bijective functions $\{f_i\}_{i=0}^{k}$, i.e., $f = f_0 \circ f_1 \circ f_2 \circ \cdots \circ f_k$. For a given observed data sample $x$, we can derive the transformation to target data sample $y$ through

$$y = f_0 \circ f_1 \circ f_2 \circ \cdots \circ f_k(x) \quad (3)$$
$$x = f_k^{-1} \circ f_{k-1}^{-1} \circ \cdots \circ f_0^{-1}(y) \quad (4)$$

The bijective model $f_i$ is implemented through affine coupling layers. In each affine coupling layer, given a $D$ dimensional input $m$ and $d < D$, the output $n$ is calculated as

$$n_{1:d} = m_{1:d} \quad (5)$$
$$n_{d+1:D} = m_{d+1:D} \odot \exp(s(m_{1:d})) + t(m_{1:d}) \quad (6)$$

where $s$ and $t$ represent scale and translation functions from $R^d \to R^{D-d}$, and $\odot$ is the Hadamard product. Note that the scale and translation functions are not necessarily invertible, and thus we realize them by neural networks.

As stated in [45], the coupling layer leaves some input channels unchanged, which greatly restricts the representation learning power of this architecture. To alleviate this problem, we firstly enhance [46] the coupling layer by

$$n_{1:d} = m_{1:d} + r(m_{d+1:D}) \quad (7)$$

where $r$ can be arbitrary function from $R^{D-d} \to R^d$. The inverse step is easily obtained by

$$m_{d+1:D} = (n_{d+1:D} - t(n_{1:D})) \odot \exp(-s(n_{1:d})) \quad (8)$$
$$m_{1:d} = n_{1:d} - r(m_{d+1:D}) \quad (9)$$

Next, we utilize the invertible 1×1 convolution proposed in [25] as the learnable permutation function to reverse the order of channels for the next affine coupling layer.

The elaborate architecture of iVAN is shown in Fig. 4. It contains several invertible blocks, where each invertible block consists of invertible 1×1 convolution and affine coupling layers. The input image is split into two halves along the channel dimension. $s$, $t$ and $r$ are transformation equal to dense block, which consists of five 2D convolution layers with filter size 3×3. Each layer learns a new set of feature maps from the previous layer. The size of the receptive field for the first four convolutional layers is 3×3, and stride is 2, followed by a rectified linear unit (ReLU). The last layer is a 3×3 convolution without ReLU. The purpose of the Leaky ReLU layers is to avoid overfitting to the training set [47] and further increase nonlinearity. Invertible model is composed of both forward and inverse process. In the forward process, the input image is transformed to output images of other modalities by a stack of bijective functions $\{f_i\}_{i=0}^{k}$.

During the training time, to invert iVAN, the backward process takes an image, which is synthetic or fused as input and inverses all the bijective functions to obtain the original image.

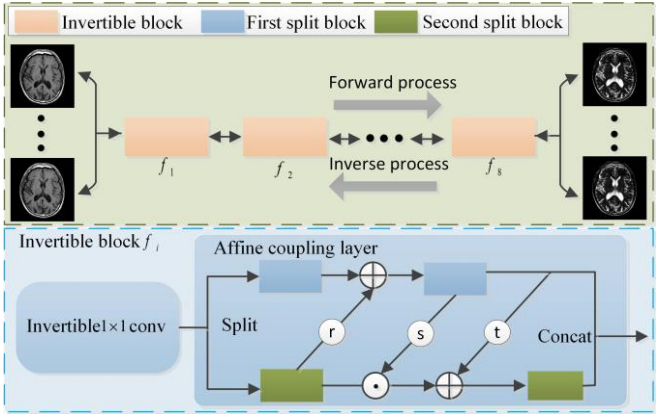

Fig. 4. The pipeline of iVAN. Invertible model is composed of both forward and inverse process. We illustrate the details of the invertible block on the bottom. r, s and t are transformations defined in the bijective functions $\{f_i\}_{i=0}^k$.

### C. Training Objectives of iVAN

In order to guarantee the quality of synthesized images, many researches leverage multi-component cost function to optimize the network. For example, in MM-Synthesis [22], a cost function constituted from three cost components is utilized to train a multi-input multi-output fully convolutional network model for MRI synthesis. In MM-GAN [20], generation loss and adversarial loss constitute the cost function of this network. The Hi-Net [21] network adds a reconstruction loss on the basis of the generation and discriminator loss. Similarly, the FGEAN [23] not only includes pixel-wise intensity loss, but also includes gradient information loss. They all use multiple loss functions to ensure the network architecture works well in a variety of synthesis tasks.

Compared with the above-mentioned methods, our loss function is simple in composition and easier to train. What is more, it can generate more faithful images than them. We use the Euclidean Loss to generate representative features. The loss function minimizes the mean squared error between pixel values of the input and the synthesis and fusion image. The training objective of iVAN is as follows:

$$\mathcal{L}_{total} = \lambda \mathcal{L}_1 + \mathcal{L}_2 \\ = \lambda \|f(X) - Y\|_2 + \|f^{-1}(Y) - X\|_2 \quad (10)$$

where $Y$ is the ground-truth target image, $f(X)$ is the output image from the source image $X$ by iVAN network $f$, and $\|\cdot\|_2$ represents the $L_2$-norm. $\mathcal{L}_1$ stands for the loss function between the synthesis image and ground-truth. $\mathcal{L}_2$ stands for the loss function between the input and reference images. Hyper-parameters $\lambda$ is used to balance the two losses.

### D. Interpretation of "Adversarial Learning" in iVAN

Recently, there existed a great number of adversarial learning approaches for image synthesis and produced promising results [20], [22], [23]. Therefore, an investigation of the relationship between these approaches and our method is worthful. In this section, we will reveal that our present iVAN network not only possesses the underlying adversarial learning scheme, but also involves compact and simple network architecture.

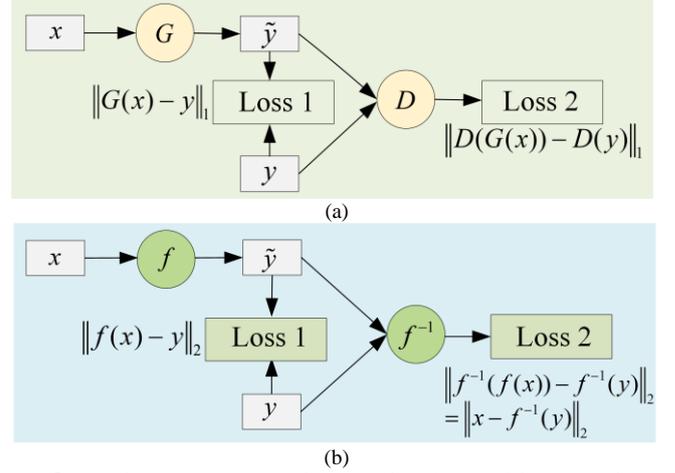

Fig. 5. Visualization comparison between the adversarial learning scheme in [14] and an interpretation of the proposed iVAN.

For the convenience of fair comparison, the method in [14] is included in the visualization comparison of Fig. 5 under unimodal synthesis. More precisely, as shown in Fig. 5(a), in many adversarial learning approaches $G(x) = \tilde{y}$, it firstly attains the synthesis image by operator $G$. Then, the transformed synthesis image $G(x)$ is compared with the reference synthesis image $y$ under the discriminative network $D$, i.e., $\|D(G(x)) - D(y)\|_1$. Hence, it often minimizes the loss functions $\|G(x) - y\|_1$ and $\|D(G(x)) - D(y)\|_1$.

By contrast, as shown in Fig. 5(b), in this work we first use the forward network $f$ to form the synthesis image $\tilde{y}$, it minimizes $\|f(x) - y\|_2$. Then, under the invertible network $f^{-1}$, we minimize the loss function $\|f^{-1}(f(x)) - f^{-1}(y)\|_2 = \|x - f^{-1}(y)\|_2$.

In summary, the proposed method iVAN attains a more compact adversarial learning mechanism, via the invertible and variable augmented techniques. notice that the objective functions in Eq. (10) are equal to the hybrid minimization of $\|f(x) - y\|_2$ and $\|x - f^{-1}(y)\|_2$.

## IV. EXPERIMENTS

In this section, we introduce the implementation details of the proposed iVAN, as well as the datasets we used for evaluation. Subsequently, the synthesis and fusion results, invertible results are reported and analyzed. Both quantitative and qualitative evaluations are comprehensively conducted to investigate the performance of iVAN.

### A. Experiment Setup

**Datasets**. We use two datasets, i.e., the Harvard university database[1] and the Brainweb dataset[2], to evaluate iVAN. To be specific, the Harvard university database contains 83 pairs of multi-modal medical images: 10 pairs of MR and CT images, 13 pairs of T1-weighted MR and T2-weighted images, 30 pairs of MR and PET images, and 30 pairs of MR and CT images. All of these source images are collected from the database of Whole Brain Atlas created by Harvard Medical School and have been widely adopted in previous publications related to medical image fusion. All the source images have the same spatial resolution of 256×256 pixels, and images in each pair have been accurately registered. The Brainweb dataset contains the full simulated brain database as well as the anatomical model used as input to the MR simulator. For each anatomical model, three imaging sequences T1, T2, PD are available online, each with a fixed set of parameters: Typical values of slice thickness, noise, and intensity non-uniformity levels. Our architecture uses 2D axial-plane slices of the volumes as the network input. For each volume, we linearly scale the original intensity values to [−1, 1]. To increase the number of training samples, we rotate the original image by different angles.

---

[1] https://www.med.harvard.edu/aanlib/.
[2] http://www.bic.mni.mcgill.ca/brainweb/.

*Model Training*. All networks are trained using the Adam solver. We conduct 300 epochs to train the proposed model. The initial learning rate is set to 0.0001 for the first 50 epochs. Every 50 epochs the learning rate is halved. During training, the trade-off parameter $\lambda$ is set to 1. The training and testing experiments are performed with a customized version of Pytorch on an Intel i7-6900K CPU and a GeForce Titan XP GPU. For the convenience of reproducible research, source code of iVAN can be downloaded from website: https://github.com/yqx7150/iVAN.

*Quality Metrics*. In the synthesis experiments, we employ three measures to evaluate the synthesis performance of the proposed iVAN model and other methods in peak signal-to-noise ratio (PSNR), structural similarity index (SSIM), and normalized mean squared error (NMSE). These three metrics are widely applied to the whole synthesized image. Denoting $\tilde{y}$ and $y$ to be the synthesized image and the ground-truth, respectively.

$$PSNR(y, \tilde{y}) = 20\log_{10} Max(y)/\|\tilde{y} - y\|_2 \quad (11)$$

SSIM is used to measure the similarity of the two images, i.e.,

$$SSIM(y, \tilde{y}) = \frac{(2\mu_y\mu_{\tilde{y}} + c_1)(2\sigma_{y\tilde{y}} + c_2)}{(\mu_y^2 + \mu_{\tilde{y}}^2 + c_1)(\sigma_y^2 + \sigma_{\tilde{y}}^2 + c_2)} \quad (12)$$

NMSE is calculated as:

$$NMSE(y, \tilde{y}) = \|y - \tilde{y}\|_2^2 / \|y\|_2^2 \quad (13)$$

For PSNR and SSIM, higher values indicate better generation of synthesis images. For NMSE, the lower the value is, the better the generation is.

In the fusion experiments, five objective fusion metrics which are commonly used in medical image fusion are adopted to make quantitative evaluations. They are the average gradient (AG), spatial frequency (SF), Entropy (EN), $Q_{MI}$, and $Q_{ab/f}$.

AG is mainly applied to assess the local contrast of an image. It is defined as the average gradient magnitude

$$AG = \frac{1}{H \times W} \sum_{h=1}^{H} \sum_{w=1}^{W} \sqrt{X_h(h,w)^2 + X_w(h,w)^2} \quad (14)$$

where $X_h$ and $X_w$ denote the gradient map of image $X$ of size $H \times W$ in $h$ and $w$ directions, respectively.

SF indicates the level of information possessed by the fused medical image.

$$SF = \sqrt{(RF)^2 + (CF)^2} \quad (15)$$

where $RF$ is row frequency while $CF$ is column frequency.

EN is mainly an objective evaluation index that measures how much information the image contains. The entropy of an image $x$ is defined as

$$H(A) = -\sum_{l=0}^{L-1} P_x(l)\log_2 P_x(l) \quad (16)$$

where $L$ is the number of gray levels and $P_x(l)$ is the normalized histogram of the image.

$Q_{MI}$ denotes normalized mutual information, which measures the amount of mutual information between the fused image and the input source image. That reflects the correlation between source images and fused images. The expression of normalized mutual information $Q_{MI}$ is as follows

$$Q_{MI} = 2[\frac{MI(x_1, \tilde{y})}{H(x_1) + H(\tilde{y})} + \frac{MI(x_2, \tilde{y})}{H(x_2) + H(\tilde{y})}] \quad (17)$$

where MI is a quantitative measure of the interdependence of two variables, $H(x_1)$, $H(x_2)$ and $H(\tilde{y})$ are information entropies.

$Q_{ab/f}$ is a quality measure based on similarity. It is defined as

$$Q(x_1 x_2 \tilde{y}) = \frac{1}{|W|} \sum_{\omega \in W} [\lambda(\omega)Q_0(x_1, \tilde{y}|\omega) + (1-\lambda(\omega))Q_0(x_2, \tilde{y}|\omega)] \quad (18)$$

where $x_1$, $x_2$ are the source image and $\tilde{y}$ is the fusion image. $\lambda(\omega)$ represents the local weight. Local Quality Index $Q_0(x_1, \tilde{y}|\omega)$ and $Q_0(x_2, \tilde{y}|\omega)$ are weighted by the local saliency that reflects the relevance with the window of an input image. For all the above five metrics, a larger score indicates a better performance.

*B. Comparisons on Image Synthesis*

To verify the effectiveness of the proposed iVAN, three state-of-the-art cross-modality synthesis methods are compared, including Pix2pix [16], cycleGAN [17], and Hi-Net [21]. These methods can be summarized as follows: 1) Pix2pix [16]. This method synthesizes a whole image by focusing on maintaining the pixel-wise intensity similarity; 2) cycleGAN [17]. This method uses a cycle consistency loss to enable training without the need for paired data. In our comparison, we use the paired data to synthesize medical images from one modality to another; and 3) Hi-Net [21]. This method uses a layer-wise fusion strategy to effectively fuse multiple modalities within different feature layers. We evaluate the proposed model for three tasks, i.e., synthesize from modality T1 to T2 on Brainweb dataset (T1→T2), synthesize from T1 and PD modalities to T2 modality on Brainweb dataset (T1+ PD → T2), and synthesize from T1 and T2 modalities to PET modality on Harvard university database (T1+ T2 → PET).

*Synthesize T2 from T1.* The quantitative evaluations for this task are recorded in Table I. It can be seen that iVAN outperforms the comparison methods, i.e., Pix2pix [16], cycleGAN [17], Hi-Net [21] in PSNR and SSIM. Specifically, iVAN gives an average PSNR of 39.4 dB, which is 2.7 dB higher than the Pix2pix method. Our method also demonstrates an overall lower standard deviation of the NMSE value throughout testing. Fig. 6 depicts the qualitative comparisons between iVAN and other state-of-the-art methods. Although all the methods produce the high-quality synthesized T2 images, the visual results generated by iVAN exhibit sharper edges, which is consistent with the observation from quantitative evaluation.

*Synthesize T2 from T1 and PD.* Our network can also be applied to the multi-input to one-output synthesis. Here, only Hi-Net [21] is included in the comparison, as it was published in the very recent 2020 and outperformed all other methods, i.e., MM-Synthesis [22], LSDN [48], and Replica [49]. The quantitative results are given in Table I. iVAN also demonstrates an overall higher PSNR and lower NMSE values compared to Hi-Net. Compared to the counterpart of one-input methods, the multi-input method shows an improvement of 3.9 dB in terms of PSNR value. In other words, using multi-input methods is indeed better than using one-input methods in our experiments. The qualitative results are shown in Fig. 7. It can be observed that iVAN is faithfully sharper, with lower blurring artifacts, while Hi-Net seems to miss most of these details.

*Synthesize PET from T1 and T2.* To further illustrate that the synthesis effect of multi-input method is better than the one-input counterpart, we set up a synthesis experiment from MRI to PET. In the experiment, we use six-channel as the input of the network. Table II tabulates the quantitative evaluation results for this task. It can be seen that multi-input method outperforms one-input version in all the three metrics. From the qualitative results in Fig. 8, the multi-input method obtains the information of the two images better and generates a reasonable color PET picture. This phenomenon indicates that our method can effectively explore complementary information from multiple modalities as well as exploit their correlations to improve the synthesis performance.

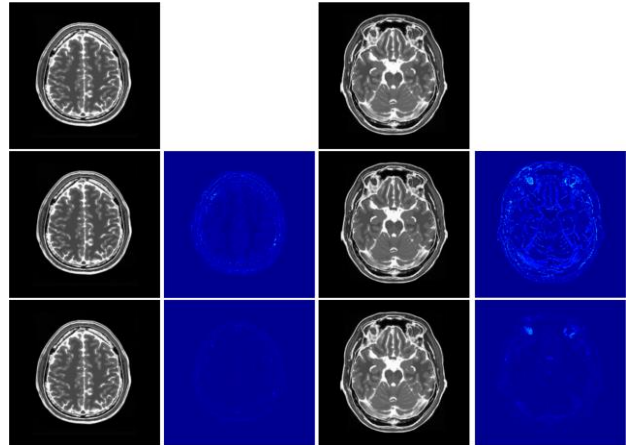

Fig. 7. Two examples of the synthesis result from T1 and PD to T2. From top to bottom are the ground-truth, the synthetic T2 by Hi-Net and iVAN, respectively. The second and fourth columns record the residuals between the synthetic T2 and the ground-truth.

TABLE I
SYNTHESIS COMPARISON WITH THREE STATE-OF-THE-ART METHODS ON BRAINWEB DATASET IN TERMS OF PSNR, SSIM, AND NMSE (MEAN $\pm$ STANDARD).

| Index | T1 $\to$ T2 | | | T1+PD $\to$ T2 | | |
|---|---|---|---|---|---|---|
| | PSNR $\uparrow$ | SSIM $\uparrow$ | NMSE $\downarrow$ | PSNR $\uparrow$ | SSIM $\uparrow$ | NMSE $\downarrow$ |
| Pix2pix | 36.6990 $\pm$ 1.3645 | 0.9926 $\pm$ 0.0022 | 0.0021 $\pm$ 0.0006 | ---- | ---- | ---- |
| cycleGAN | 35.6997 $\pm$ 3.5396 | 0.9921 $\pm$ 0.0032 | 0.0019 $\pm$ 0.0011 | ---- | ---- | ---- |
| Hi-Net | 35.6482 $\pm$ 1.6665 | 0.9911 $\pm$ 0.0070 | 0.0002 $\pm$ 0.0001 | 40.5888 $\pm$ 1.8574 | 0.9981 $\pm$ 0.0006 | 0.0008 $\pm$ 0.0002 |
| iVAN | 39.4183 $\pm$ 2.2150 | 0.9935 $\pm$ 0.0012 | 0.0012 $\pm$ 0.0004 | 43.3214 $\pm$ 1.8446 | 0.9970 $\pm$ 0.0008 | 0.0004 $\pm$ 0.0001 |

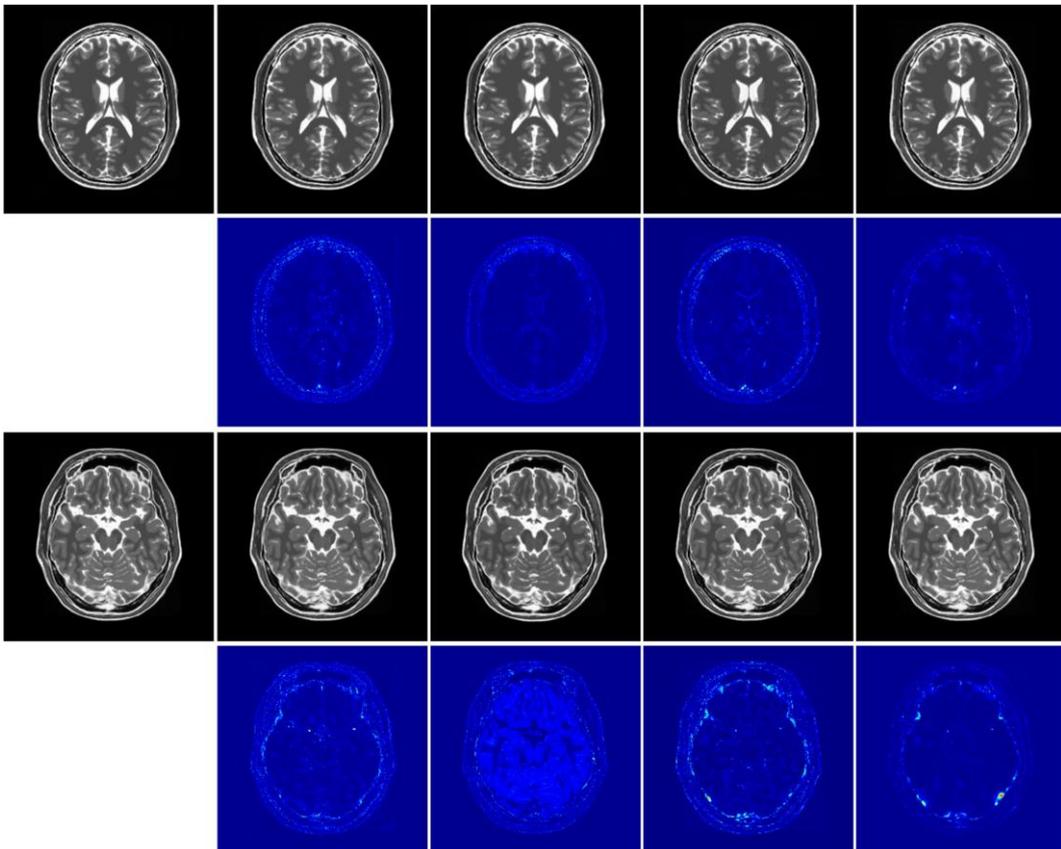

Fig. 6. Two visualization results of synthesizing from T1 to T2. The first and third rows are the ground-truth, the synthesis results of Pix2pix, cycleGAN, Hi-Net, and iVAN, respectively. The second and fourth rows depict the residuals between the synthetic T2 and the ground-truth.

TABLE II
OBJECTIVE ASSESSMENT OF IVAN ON THE HARVARD UNIVERSITY DATA-
BASE IN TERMS OF PSNR, SSIM, AND NMSE (MEAN ± STANDARD).

| Index | T1 → PET | T1+T2 → PET |
|---|---|---|
| PSNR ↑ | 21.6115 ± 1.9022 | 23.2171 ± 1.9377 |
| SSIM ↑ | 0.7047 ± 0.0511 | 0.7258 ± 0.0305 |
| NMSE ↓ | 0.1073 ± 0.0487 | 0.0723 ± 0.0190 |

## C. Comparisons on Image Fusion

The proposed network iVAN is compared with other four recently proposed medical image fusion methods, LP-SR [50], LRD [39], CS-MCA [51], and LP-CNN [35]. The parameters of these four methods are all set to the default values reported in the related publications.

Fig. 9 shows the objective performance of different methods. It can be seen that LP-CNN loses a large amount of energy, leading to a significant decrease in the intensity and contrast in many regions. Some structural details are obscured by functional information which degrades the visual perception to some extent. The method LP-SR generally performs well, but intensity inconsistency exists in some regions. The LRD method can preserve the image energy, but fails in extracting the structural details from the MR image, and it can be observed that many small details of the source images are blurred in the fused images. CS-MAC performs better than the method LRD, while some details are still not successfully extracted. The proposed fusion method generally does well in both detail and energy preservation without introducing undesirable visual artifacts.

The objective assessment of different fusion methods is listed in Table III. For each metric, the average score of each method over twenty-two source image pairs is provided and the value exhibit as bold denotes the best score among all the methods. iVAN outperforms other four methods on all the metrics except for EN index which reflects that the images obtained by our method are better than all competing methods in terms of similarity and clarity. Moreover, the advantages of iVAN are generally outstanding.

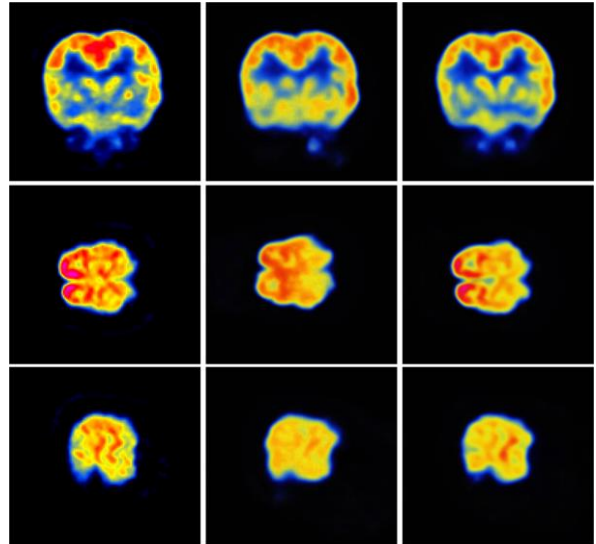

Fig. 8. Qualitative results for the PET synthesis task on Harvard university database. From left to right are ground-truth, one-input, and multi-input synthesis results.

TABLE III
COMPARED TO FOUR DIFFERENT METHODS ON BRAINWEB DATASET IN TERMS OF AG, SF, EN, $Q_{MI}$, AND $Q_{ab/f}$ VALUES.

| Method | AG ↑ | SF ↑ | EN ↑ | $Q_{MI}$ ↑ | $Q_{ab/f}$ ↑ |
|---|---|---|---|---|---|
| LP-SR | 9.3826 | 36.9809 | 4.2533 | 0.6984 | 0.5930 |
| LRD | 9.3319 | 34.2692 | 5.0014 | 0.7016 | 0.4800 |
| CS-MCA | 8.2714 | 35.0354 | 4.5341 | 0.6746 | 0.5760 |
| LP-CNN | 6.5648 | 24.5249 | **5.9304** | 0.5919 | 0.5002 |
| iVAN | **11.3779** | **39.0257** | 4.8530 | **0.7094** | **0.5937** |

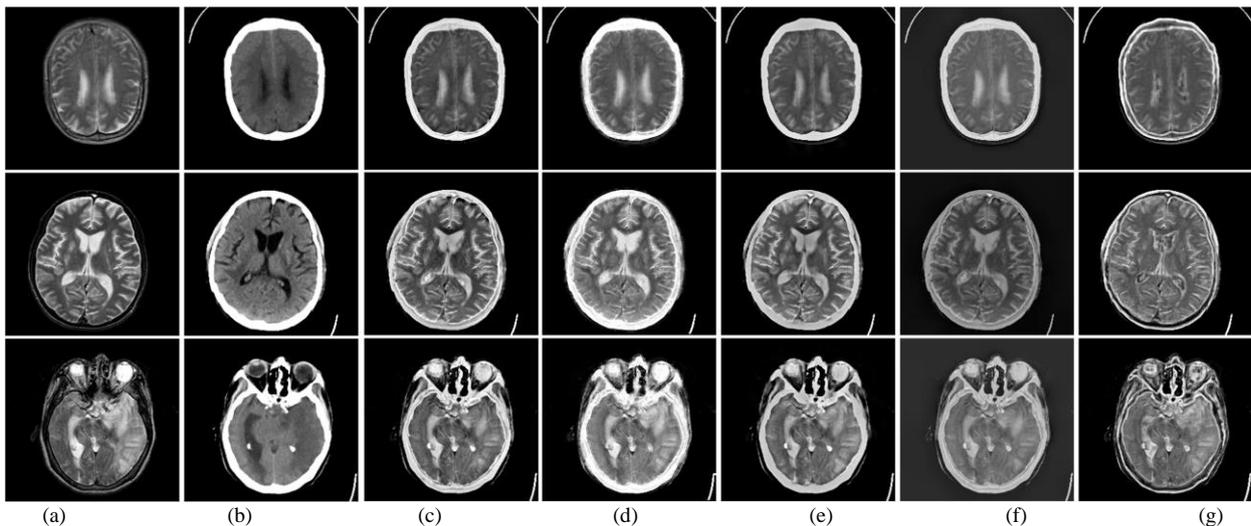

(a) (b) (c) (d) (e) (f) (g)

Fig. 9. Three fusion results of T2-weighted MR and CT images. From left to right: (a) T2-weighted MR source images, (b) CT source images, (c) LP-SR based method, (d) LRD based method, (e) CS-MCA based method, (f) LP-CNN based method, (g) iVAN.

## D. Invertible Image Synthesis and Fusion

To better verify the invertibility of iVAN, we display the invertible results in Fig. 10. It can be seen that even the inverse process can recover the true and diverse original images. iVAN achieves the best performance in terms of noise-artifact suppression and tissue feature preservation. Besides, the visual appearance of the invertible results is very similar to the ground-truth image. The quantitative results achieved by synthesis and fusion are presented in Table IV, in terms of two different evaluation metrics. iVAN has achieved very high values on all indicators. Specifically, it can reach an average of 62 dB on PSNR value.

TABLE IV
OBJECTIVE ASSESSMENT OF iVAN FOR INVERTIBLE IMAGE SYNTHESIS AND FUSION IN TERMS OF PSNR AND SSIM (MEAN $\pm$ STANDARD).

| Index | T1+PD ⟷ T2 | | T1+T2 ⟷ PET | | T2+CT ⟷ Fusion | |
|---|---|---|---|---|---|---|
| | T2 → T1 | T2 → PD | PET → T1 | PET → T2 | Fusion → T2 | Fusion → CT |
| PSNR ↑ | 56.0053 ± 4.1925 | 58.0174 ± 2.0429 | 66.0272 ± 10.7341 | 68.9461 ± 10.9997 | 59.8782 ± 6.3086 | 64.8058 ± 4.1504 |
| SSIM ↑ | 0.9993 ± 0.0004 | 0.9969 ± 0.0015 | 0.9997 ± 0.0002 | 0.9995 ± 0.0006 | 0.9978 ± 0.0023 | 0.9996 ± 0.0002 |

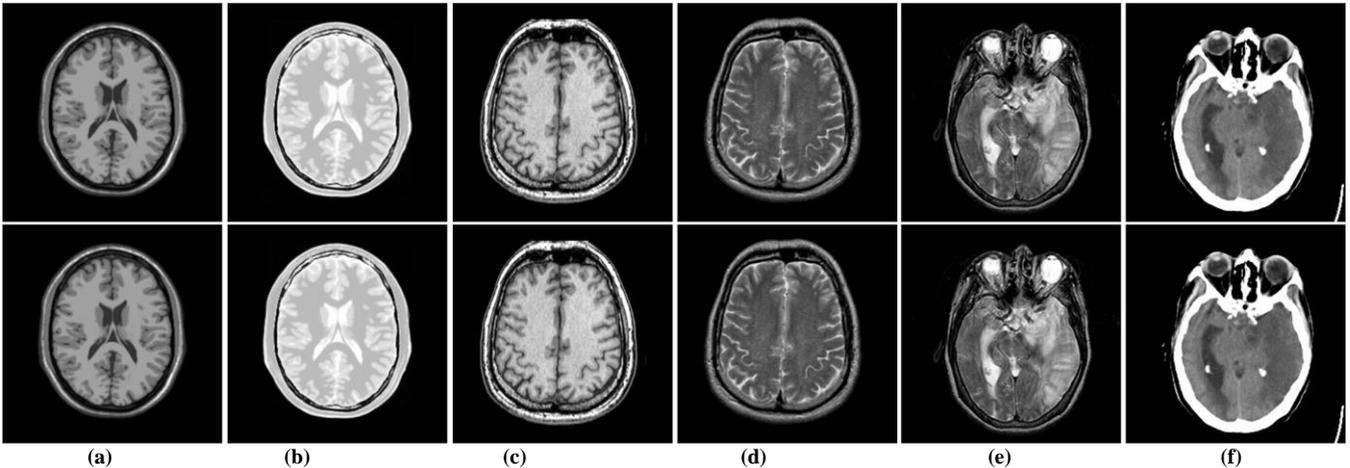

(a)    (b)    (c)    (d)    (e)    (f)

Fig. 10. Inverse result of image synthesis and fusion. The first row exhibits the source images. In the second row, from left to right: (a) T2 → T1 (b) T2 → PD (c) PET → T1 (d) PET → T2 (e) Fusion → T2 (f) Fusion → CT.

## V. DISCUSSION

In this section, we bring attentions to the choice of loss functions in iVAN. Specifically, L2-norm penalizes larger errors, but it is more tolerant to small errors regardless of the underlying structure in the image. Compared to L2-norm, L1-norm does not over-penalize larger errors. Consequently, they may have different convergence properties at the network training procedure. Inspired by this observation, we compare the synthesis effect of the loss function equipped with the L1-norm and L2-norm, respectively.

TABLE V
THE IMPACT OF LOSS FUNCTIONS ON iVAN (MEAN $\pm$ STANDARD).

| Index | T1+PD → T2 | |
|---|---|---|
| | L1-norm | L2-norm |
| PSNR ↑ | 42.0547 ± 2.0842 | 43.3214 ± 1.8446 |
| SSIM ↑ | 0.9984 ± 0.0005 | 0.9970 ± 0.0008 |
| NMSE ↓ | 0.0006 ± 0.0003 | 0.0004 ± 0.0001 |

Empirical results in Table V indicate that the loss function constrained by the L2-norm achieves higher PSNR and lower NMSE values. Specifically, the average PSNR value of the L2-norm constrained loss function is 1.3 dB higher than the synthesis effect of the L1-norm constrained loss function. At the same time, L2-norm as a mainstream constraint function also has its advantages, mainly because it is convex and differentiable, it is very convenient for optimization problems. Therefore, the L2-norm constrained loss function is selected in the experiment to optimize the learnable parameters in iVAN network.

As well known, multi-channel can enhance the correlation between data. To further investigate the effectiveness of the multi-channel mechanism, we conduct comparisons of the synthesis between that obtained by three-channel and six-channel objects in iVAN. Table VI lists the performance of iVAN under different channels.

TABLE VI
THE IMPACT OF SIX-CHANNEL AND THREE-CHANNEL ON iVAN SYNTHESIZE PET FROM T1 AND T2 (MEAN $\pm$ STANDARD).

| Channel | 3-channel | 6-channel |
|---|---|---|
| PSNR ↑ | 22.0593 ± 1.6853 | 23.2171 ± 1.9377 |
| SSIM ↑ | 0.6915 ± 0.0359 | 0.7258 ± 0.0305 |
| NMSE ↓ | 0.0943 ± 0.0344 | 0.0723 ± 0.019 |

The observation in Table VI illustrates that using six-channel as the input of the network obtains a better synthesis effect. Because the high-dimensional structures can represent distinct but complementary information that exists at various scales. The correlation between input data is enhanced. Thus, we choose six-channel as the input for the synthetic PET image to obtain better synthetic performance.

## VI. CONCLUSION

In this work, invertible synthesis and fusion were introduced on the basis of the invertible neural network. Specifically, iVAN explored the modality-specific properties within each modality, and simultaneously exploited the correlations across multiple modalities. Variable augmentation technology was enforced on the invertible network to make the network input and output to be the same dimension, thereby treating image synthesis and fusion as a unified framework. In addition, the input and output of the network are both involved in the design of the loss function, thus realizing many-to-many mapping with a simple network structure. Experimental results in synthesis and fusion tasks have demonstrated that our proposed model outperformed other state-of-the-art methods in both quantitative and qualitative measures. In forthcoming studies, we will further investigate the potential of the invertible mechanism for accelerated multi-modal imaging, built on the multi-modal synthesis and fusion.